# Structured illumination microscopy with extreme ultraviolet pulses

R. Mincigrucci[1], E. Paltanin[1,2], J.-S. Pelli-Cresi[1], F. Gala[3,4], E. Pontecorvo[3,4], L. Foglia[1], D. De Angelis[1], D. Fainozzi[1], A. Gessini[1], D. S. P. Molina[5,6], O. Stranik[7], F. Wechsler[7,8], R. Heintzmann[7,9], G. Ruocco[3,10], F. Bencivenga[1], C. Masciovecchio[1]

1) Elettra Sincrotrone Trieste SCpA, Strada Statale 14 - km 163,5 34149 Basovizza, Trieste ITALY
2) Department of Physics, Università degli Studi di Trieste, 34127 Trieste, Italy
3) Istituto Italiano di Tecnologia, Viale Regina Elena 291, 00161, Rome, Italy
4) CRESTOPTICS SPA Via di Torre Rossa 66, 00165, Rome, Italy
5) Institute of Applied Physics, Friedrich-Schiller-Universität Jena, 07745 Jena, Germany
6) Helmholtz-Institute Jena, Fröbelstieg 3, 07743 Jena, Germany
7) Leibniz Institute of Photonic Technology, Albert-Einstein Straße 9, 07745 Jena, Germany
8) Laboratory of Applied Photonics Devices, Ecole Polytechnique Fédérale de Lausanne (EPFL), Switzerland
9) Institute of Physical Chemistry and Abbe Center of Photonics, Helmholtzweg 4, 07743 Jena, Germany
10) Department of Physics, Sapienza University, Piazzale Aldo Moro 5, 00185, Rome, Italy

**Abstract**
The relentless pursuit of understanding matter at ever-finer scales has pushed optical microscopy to surpass the diffraction limit and produced the super-resolution microscopy which enables visualizing structures shorter than the wavelength of light. In the present work, we harnessed extreme ultraviolet beams to create a sub-$\mu m$ grating structure, which was revealed by extreme ultraviolet structured illumination microscopy. This achievement marks the first step toward extending such a super-resolution technique into the X-ray regime, where achieving atomic-scale resolution becomes a charming possibility.

## 1- Introduction

During the 17th century, a significant advancement occurred with the invention of the first optical instruments, which expanded our vision from distant planets to the intricate world of small objects surrounding us. This initiated the race to achieve higher magnifications, allowing scientists to delve into the finer details of matter. Over time, microscopy evolved into a distinct scientific field, going far beyond what the human eye can perceive, and it routinely employs photons beyond our visual range.

A major breakthrough in optical microscopy was made by Abbe, who developed the fundamental theory of image formation in microscopes (1) and their limitations. Indeed, the smallest pitch a periodic structure must have to be resolved in a microscope, $d$, is defined through the Abbe relation as $d = \frac{\lambda}{2n \sin\theta}$, where $\lambda$ is the vacuum wavelength of the light used to obliquely illuminate the specimen, $n$ is the refractive index of the immersion medium between the sample and the objective and $\theta$ is the maximum semi-angle collected by the imaging optics. The wavelength is the only limitless parameter in the Abbe's equation and for decades the privileged way to increase the resolution, *i.e.,* to decrease $d$, was to use shorter values of $\lambda$, which led to the realization of X-ray microscopes. Exploiting the X-ray radiation produced by synchrotrons and free electron lasers (FELs), several imaging techniques have been implemented during the years, reaching $d \approx$ 20 nm (2,3). However, such a resolution is still significantly larger than the X-ray wavelength ($\lambda <$ 1 nm), with the lower bound to the resolution given by the manufacturing of efficient X-ray optics.

Alternative approaches, based on electrons for illuminating the sample (4,5) or on proximity forces (6) have been developed and effectively utilized (7,8). However, these techniques entail some burdens: for instance, atomic force microscopes provide an atomic-scale resolution but are only capable of imaging the very top surface layer, while electron microscopes can reach sub 2 Å resolution (9), however, need ultra-high vacuum and sub-micrometer thick samples to obtain an exploitable electron mean-free-path.

Using short wavelengths is not the only way for increasing the resolution, a different approach is to challenge the Abbe equation. The work of G. Toraldo Di Francia firstly questioned the resolution limit (10) introducing the concept of super-resolution and discussing the possibility of obtaining it using visible light. Indeed, super-resolved techniques have revolutionized microscopy (11–13) and, among them, a conceptually simple one is fluorescence structured illumination microscopy (SIM).

In SIM, a spatially periodic pattern of light intensity is used to illuminate the sample. If the spatial frequency of the light pattern ($k_i$) is similar to a characteristic spatial frequency of the sample ($k_s$), the emitted fluorescence exhibits a pattern at the frequency $k_s - k_i$, known as moiré fringes, which can often be imaged, i.e. transferred by the imaging system. The corresponding grating pitch of such pattern is $L_M \sim 2\pi/(k_s - k_i)$, which can be larger than the individual characteristic lengths $L_s$ and $L_i$, associated, respectively, to $k_s$ and $k_i$. Therefore, even if a feature of the sample is smaller than the aforementioned Abbe limit $d$, it can be detected via its downmodulation by structured illumination, which can result into a detected pattern with $L_M > d$. Thus, SIM makes Abbe-forbidden features accessible by generating patterns that can be transferred (imaged) by the microscope (14–16).

State-of-the-art microscopes based on visible light are able to access the full range of spatial information allowed by the Abbe limit, therefore, the use of visible radiation to generate the structured illumination can extend the spatial resolution by an amount equal to $k_i$. This is shown in Figure 1a, where the empty black circle represents the range accessible by standard (Abbe limited) fluorescence microscopy, while the yellow circles are translated by $|k_i|$. By capturing multiple images with different orientations and phases of the objective-launched illuminating patterns, standard SIM allows to retrieve the sample structure with a resolution about twice beyond the Abbe limit (17) represented in Figure 1b as the yellow dashed circle. Here it is important to note that an overlap between the Abbe limited resolution circle and the translated ones is considered necessary in the reconstruction process. Indeed, the overlap area is required to match, in terms of contrast and positioning in the parent image, details present both in the standard and extended resolution circles (14).

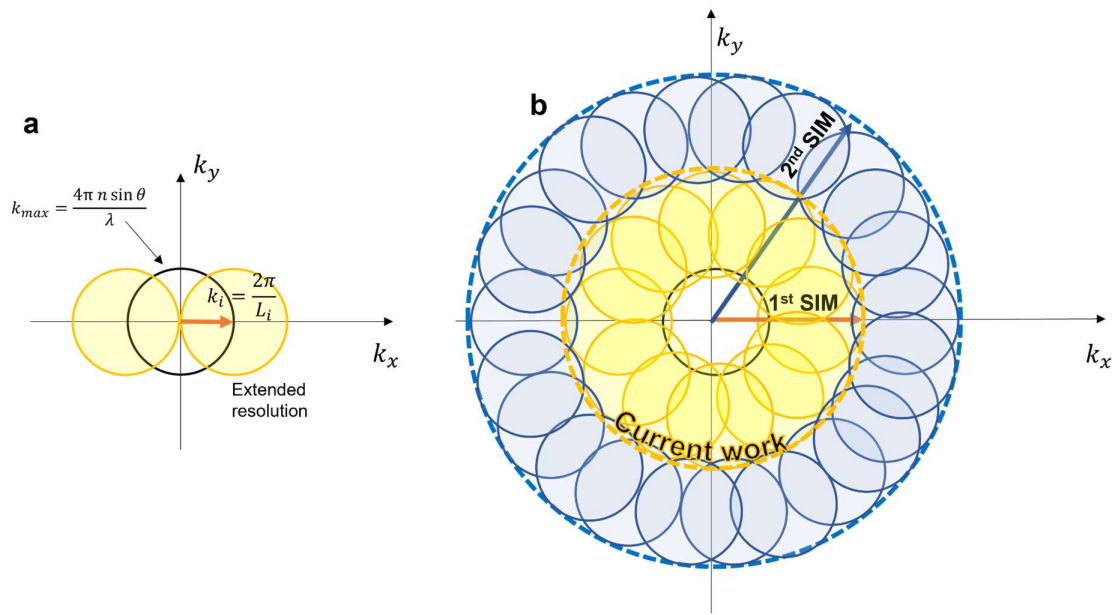

Figure 1: Panel a) sketch of standard SIM microscopy. The black circle encloses the spatial frequencies $k_s < k_{max} \approx 2\pi/d$ accessible by an Abbe-limited optical microscope, while yellow circles are those accessible by using a structured illumination with characteristic spatial frequency $k_i$. Using visible light the value of $k_i$ is in the same order as $k_{max}$. Panel b illustrates the possibility to use EUV TG to progressively increase the accessible range in $k_i$ while maintaining overlap in $(k_x, k_y)$-space to not leave any gaps.

In the present work, we explored the charming possibility of extending the SIM technique not only beyond the visible spectral range (18) but also in the time domain. Indeed, we used extreme ultraviolet EUV light pulses to generate the structured illumination causing visible fluorescence on a scintillator crystal. The value of $k_i$ can be continuously varied from $k_{max}$ up to the inverse nanometer range, in $k_i$-steps sufficiently small to preserve the overlap in the $(k_x, k_y)$-space, as shown in Figure 1b. Indeed, the resolution extension process can be carried on iteratively, as sketched in Fig 1b, where the new translated circles (blue ones) have to overlap with the 1st SIM one, which is already beyond the Abbe limit. The iterative extension of the resolution will be the object of future work.

Specifically, we used the special setup realized for EUV transient grating (TG) experiments (19–21) for generating an EUV structured illumination with nanoscale spatial periodicity. The latter, combined with the wavelength flexibility offered by the FERMI FEL, was used to generate a grating with period $L_s$ and to illuminate it with a similar grating of period $L_i$. This generates coarse moiré fringes in the optical fluoresence that can be imaged and used to reconstruct a super-resolved image with almost three times the original resolution.

## 2 - Results

In an EUV TG experiment (20) two parallelly polarized EUV beams are crossed to generate a sinusoidal modulation of light intensity with period $L_{TG} = L_i = \lambda/(2 \sin \theta)$, where $\lambda$ is the EUV wavelength and $\theta$ is the half angle between the two beams; see the Methods section for further details. Moiré patterns can have a complex shape, depending on the shape of the sample; the simplest moiré pattern is the one given by a sinusoidal sample. Such "sinusoidal samples" can be generated "on-site" using EUV TGs at fluences above the damage threshold that permits a periodical modulation of the fluorescence efficiency. This is likely caused by the generation of color centers where the FEL intensity exceeds the damage threshold (22–24). The fine control over $\lambda$ offered by the FERMI FEL, where the EUV TG instrument is located, permits to control the expected moiré periodicity. Indeed, after generating a sample with a given value of $L_s$ using a high fluence structured beam, the same sample can be illuminated by a (lower fluence) structured beam with a periodicity $L_i \neq L_s$, so that the value of $L_M$ can be varied and brought in the few $\mu m$ range for the fluorescence to be observable by a visible microscope. By using a fluence of 1 mJ/cm² at $\lambda = 83.2$ nm, we generated a permanent sinusoidal spatial modulation of fluorescence efficiency with period $L_s = 263$ nm in a Cerium doped Yttrium Aluminum Garnet (YAG) crystal flake of 20 $\mu m$ thickness, which is an efficient scintillator for EUV radiation. The YAG sample was then illuminated by a EUV TG at much lower flux (in order to not permanently modify the sample anymore) and the resulting fluorescence was collected using an objective with theoretical resolution of $d = 0.37 \, \mu m$. See methods section for further details. The periodicity of the modulated fluorescence efficiency is thus well beyond this limit and is not distinguishable for our microscope. To achieve the superresolution with FEL pulses, we acquired 250 images by illuminating the sample both with EUV TG (structured illumination) and with a standard FEL pulse (homogeneous illumination), obtained by shutting one of the pump beams. Fluence was set to 0.08 mJ/cm$^2$ and $\lambda$ to 86.6 nm to generate an EUV TG with $L_i = 274$ nm. Considering that both the modulation of the sample's fluorescence efficiency and the structured EUV beam are sinusoidal, the expected moiré pattern is also sinusoidal, with fringe spacing of $L_M$= 6.6 $\mu m$, which falls well within the resolution of our microscope; i.e. $L_M > d$. GIF files demonstrating the appearance of such fringes when the structured illumination generated by the FEL pulses is directed onto the sample are included in the supplementary material. Due to small and unavoidable mechanical instabilities of the TIMER beamline, the phase and frequency of the illuminating beam are subject to random fluctuations which prevents the use of standard reconstruction algorithms. Figure 2a shows the image of the YAG crystal obtained by using homogeneous illumination, while Figure 2b displays the reconstructed image, as obtained through structured illumination. The latter was achieved by a Bayesian approach based on a joint likelihood minimization over the 250 images. Both images have been registered to cancel the stochastic sample motion. See methods for further details. An improvement in the image contrast (edge sharpness) is evident already by visual inspection. In addition, in Figure 2b a fine grating structure can be perceived, which is confirmed by the projections of the colored boxes in the bottom right corners. These are depicted in Figure 2c, where the red trace represents the projection of the image obtained with homogeneous illumination, and the green one represents the projection of the image reconstructed after structured illumination. A linear background has been subtracted from both traces. The permanent sinusoidal modulation of the fluorescence efficiency, invisible in the case of homogeneous illumination, appears in the green trace. To better appreciate the resolution gain, we display in Figure 2d the unidimensional Fourier transform of the homogeneously illuminated sample (red trace) and of the reconstructed image (green trace), compared with the Optical Transfer Function (OTF, black dashed line). The latter limits the resolution in standard microscopy and ultimately defines the smallest details that are transferred to the observer with an amplitude greater than zero. The trace corresponding to homogeneous illumination is contained in the OTF which reaches the zero value at about 2.5 $\mu m^{-1}$ which thus impedes the detection of structures smaller than approximately 400 nm. On the contrary, a peak at about 3.8 $\mu m^{-1}$, well outside the OTF, appears on the Fourier transform of the reconstructed image. This evidences the increased resolution, which now reached about 6 $\mu m^{-1}$ (limit of the OTF translated by $1/L_i$, dash-dotted black line) and permits to visualize otherwise hidden structures. Remarkably, the resolution is increased by a factor 2.4 greater than the one usually achieved in the optical domain SIM.

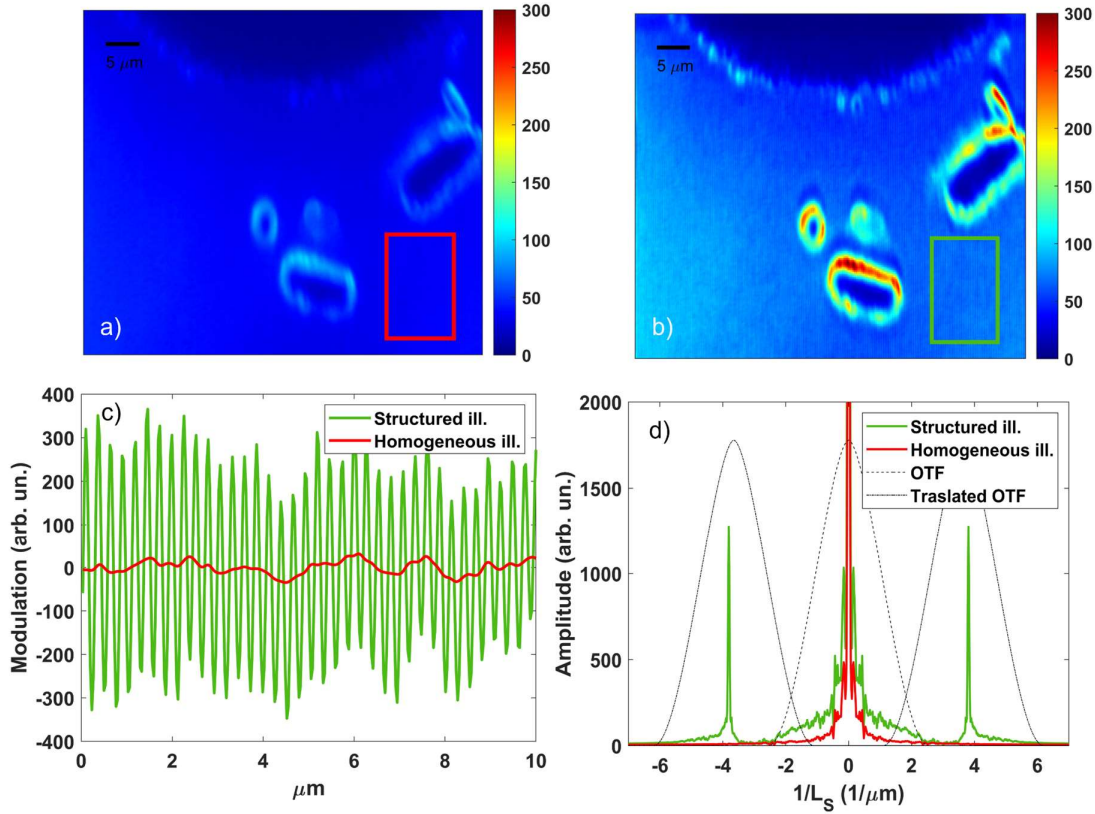

Figure 2- Panel a) shows the image of the patterned YAG when illuminated with a homogeneous beam. Panel b) displays the same portion of the YAG after the reconstruction. In panel c) are plotted the projections of the colored boxes in panel a) and b); red and green traces correspond to homogeneous and structured illumination, respectively. In panel d) are shown the unidimensional Fourier transform in the direction of the modulation of the original image, obtained with homogeneous (red trace) and structured (green trace) illumination, as well as the optical transfer function of the system (black dashed curve) and the translated one (black dash-dotted curve).

## 3 - Discussion

In this experiment we used EUV TG to demonstrate EUV structured illumination microscopy.

A complete reconstruction of the structure would have required rotating the sample with respect to axis of the structured beam, thus spinning the extended resolution circles (blue circles in Fig. 1a) all over the $k_{xy}$ plane. This is easily implementable, however, in this particular case, it is not necessary since the sample has a unidimensional structure, with the only relevant $k$ component residing on the $k_x$ axis. With the current EUV TG capabilities, SIM with $L_i$ as short as a few 10s of nm can already be envisioned, opening the possibility of extending this linear SIM super-resolution technique to the sub-100 nm or even to the single-digit nm regime, i.e. to reach the dimension of large molecules, as sketched in Figure 3. In particular, the continuous tunability in $\lambda$ provided by the FERMI FEL allows a continuous variation of the extended resolution circles up to about 2 nm, with an arbitrarily large degree of overlap among the resolution circles.

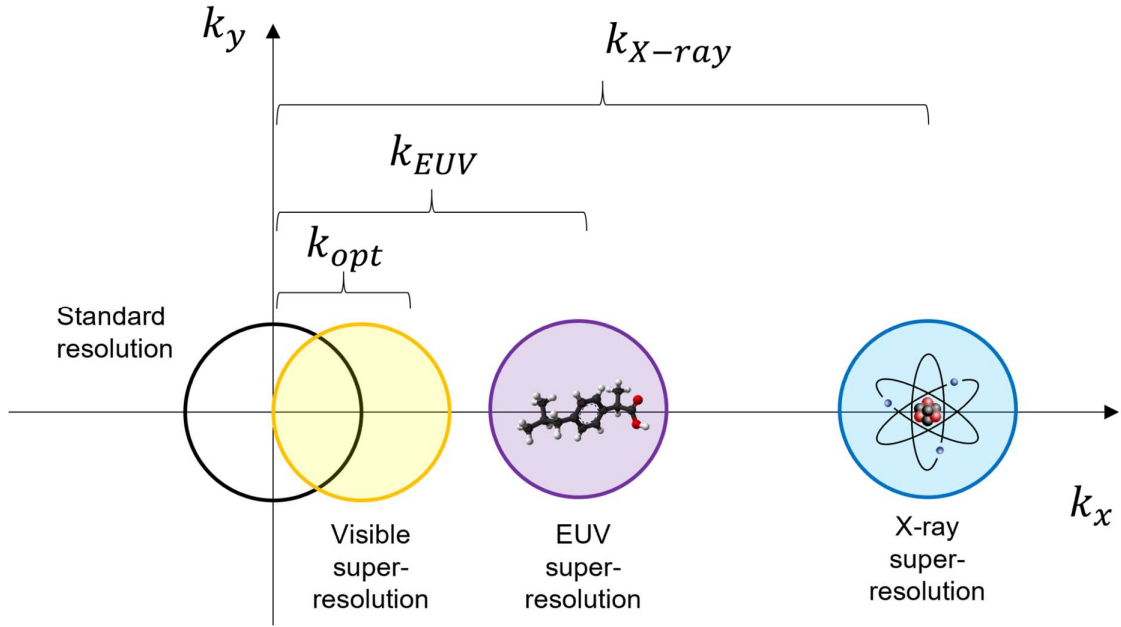

Figure 3 - Sketch of the potentiality of SIM beyond the visible spectral range. Available EUV TG instruments might reach the dimensions of a large molecule, X-ray interferometers can generate sub-nm X-ray structured beams, potentially enabling the visualization of atoms.

To go beyond the current capabilities of EUV TGs in terms of periodicities reached by TG, the use of X-ray structured beams is needed. To generate such beams we envision using Laue or Bragg interferometers (25); there nm and sub-nm periodicities can be achieved and varied by choosing the crystal material and reflection order.

A key aspect of X-rays with respect to visible light is the possibility to exploit X-ray fluorescence, which possesses two intrinsic advantages: i) it is element specific and ii) is emitted in the fs scale (26). The latter aspect can straightforwardly enable visualizing the sample dynamics atom-wise on ultrafast timescales via pump-probe approaches. Moreover, the former aspect can be used to obtain chemical sample mapping and obtain relevant information for a variety of fields which ranges from medicine (27) to cultural heritage (28).
The feasibility of X-ray fluorescence imaging has already been demonstrated (29) and the high brightness of modern X-ray FELs makes single-shot capabilities envisionable. Indeed, considering a 100 x 100 $\mu m^2$ spot, $10^{12}$ photons/pulse and a 100-$\mu m$ thick solid-state sample, one can realistically expect a few thousands of photons emitted over $4\pi$.

The potential applications of X-ray illuminated SIM (XSIM) are vast and particularly apply to disordered systems, where diffraction techniques can be used to visualize only the average structures of matter. For example, XSIM can be used to study batteries(30) that power our vehicles, devices and houses, where, *e.g.*, the lithium dislocations may be observed during and after a fast charging, or to study absorption and desorption dynamics on catalytic surfaces and nanoparticles(31–34).

In conclusion, we used crossed EUV pulses to generate "on site" a sub-$\mu m$ grating and observe it through structured illumination microscopy. Our approach for EUV SIM already permitted us to double the spatial resolution of single beam, synchrotron-based X-ray SIM(18) and for the first time was obtained with a pulsed source. Indeed, the natural evolution of the technique regards the use of X-ray structured beams with single-digit nanometer periodicities and beyond, and X-ray fluorescence imaging. This could be the key step towards studying both structure and dynamics of the sample, atom-by-atom with possibilities currently beyond imagination. To understand how matter works, we need to see it.

# 4 - Methods

**4.1 - Experimental set-up -** To demonstrate EUV SIM we used our EUV TG setup TIMER(20) in combination with a fluorescence microscope with a theoretical spatial resolution $d = 0.37 \ \mu m$ (Zeiss Epiplan 50x, NA = 0.7, infinity corrected, located in vacuum) and capability to collect scintillation fluorescence images from a single FEL shot illumination. The objective was mounted on a 5 axis degree of freedom manipulator, which permitted adjusting the objective to sample distance, its transversal position with respect to the FEL beams and two rotations. The set-up was completed by a tube lens outside the vacuum window imaging on an ORCA sCMOS camera from Hamamatsu, both located in air. See Figure 4. We used the EUV TG setup to generate sinusoidal structured illumination patterns with spatial periods determined by the grating equation: $L_i = \lambda/(2 \sin \theta)$. In the current experiment, $\theta$ was set to $9.1°$ while $\lambda$ was varied in order to change $L_i$. As a sample we used a 20 $\mu m$ thick YAG crystal, which is a known and efficient scintillator material for converting EUV light into visible light at about 550 nm. The sample was purposely damaged using $\lambda = 83.2$ nm and an FEL fluence of about 1 $mJ/cm^2$ to imprint on the YAG surface a permanent sinusoidal modulation of period $L_s = 263$ nm. The light emitted by the YAG was collected by our objective, but the $L_s$ modulation falls well beyond $d$ and any modulation can be observed when the sample is illuminated by a flat (not-structured) beam. In order to reveal such a modulation via structured illumination, a low fluence EUV TG with period $L_i = 273$ nm was obtained by tuning $\lambda$ to 86.6 nm and by reducing the FEL fluence at the sample position down to about 0.08 $mJ/cm^2$, well below the damage threshold. Since both the structure to be visualized and the illuminating one are sinusoidal gratings, the resulting interference pattern is also a sinusoidal grating but with a much longer period of $L_M \sim 6.6 \ \mu m$, which falls well within the resolution of the employed fluorescence microscope. The moiré fringes can thus be easily imaged as can be seen in the attached GIF file.

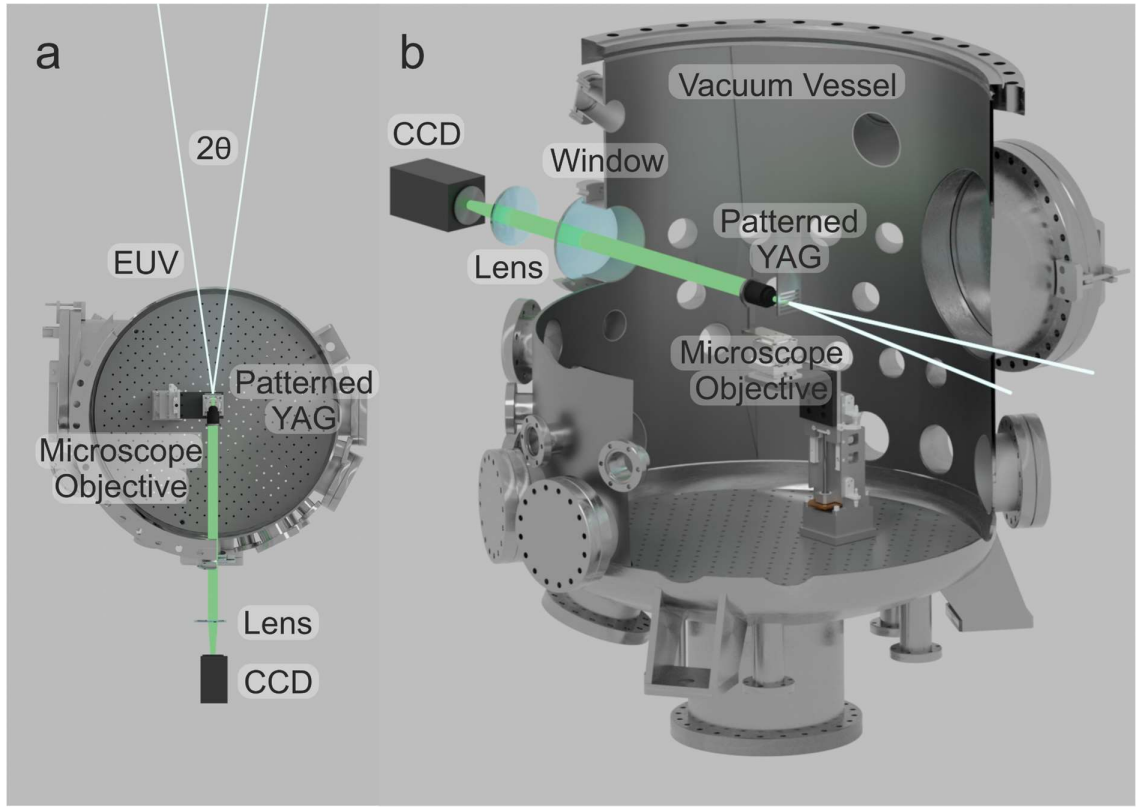

Figure 4 - View of the experimental set-up from above (panel a) and from the side (panel b). White rays are the EUV beams that cross at the sample position (i.e. the patterned YAG) with an angle $\theta = 9°$. The fluorescence light, which is spectrally peaked at 550 nm wavelength, was collected through the microscope objective, steered outside the vacuum vessel of the experimental chamber and focussed by a lens at the camera (CCD) position. The objective was mounted on a 5-axis manipulator.

**4.3 - Image reconstruction -** In standard SIM, the reconstruction of the super-resolved (SR) image is based on the precise knowledge of the frequency and phase of the illuminating grating, which allows to reconstruct frequency $(\frac{2\pi}{L_s})$ and phase $(\varphi_s)$ of the structure under examination. In the current case, the frequency and phase of the illuminating beam were fluctuating on a shot-to-shot basis due to the mechanical instabilities of the beamline and the standard reconstruction algorithm cannot be applied(15,35). For this reason, we employed a statistical Bayesian approach, based on likelihood minimization (35). Indeed, $\varphi_s \in [0, 2\pi)$ can be retrieved as the value that minimizes the joint likelihood between all the collected data and a model fluorescence image constructed with the j-th sinusoidal illumination pattern : $m_j(x,y) = 0.5 * (1 + \cos(\frac{2\pi}{L_i^{(j)}} x + \varphi_i^{(j)}))$ which results in a moiré pattern of frequency $(k_s - k_i^{(j)})$ and phase $(\varphi_s - \varphi_i^{(j)})$. The reconstruction of the unknown image has then been achieved with the following procedure: i) all the images have been registered using a monomodal, intensity-based, algorithm from the Matlab Image processing Toolbox. The algorithm has been applied to the main fixed feature (black box in Fig. 5a) to retrieve the translation which was then applied to the entire image. This approach minimizes the effect of the moiré fringes which can make the correlation-based registration fail; ii) a sinusoidal fit was performed on a featureless image portion to determine the frequency and phase of the moiré fringes. Starting parameters of the fit have been obtained by a discrete Fourier Transform; iii) the image x-axis has been enlarged to contain 4 times more pixels and the pixel values have been calculated using a bilinear interpolation; iv) a value $\varphi_s$ has been extracted from $[0,2\pi)$ interval and used to calculate the expected interference pattern with which; v) an iterative procedure based on a joint Richardson-Lucy (36,37) deconvolution has been applied to reconstruct the SR image. In the latter, the probability distribution for the image intensities has been chosen as a sum of a Poissonian distribution representing the intrinsic intensity fluctuations of the FEL, and of a Gaussian one which accounts for the camera read-out noise. Ten iterations for each value of $\varphi_s$ have been performed to reach a stable value. The results of likelihood estimation are shown in Figure 5b. The procedure has been tested on the whole image and on the region of interest evidenced in Figure 5 a) containing in both cases $\varphi_s \sim \frac{3\pi}{2}$, validating the robustness of the approach.

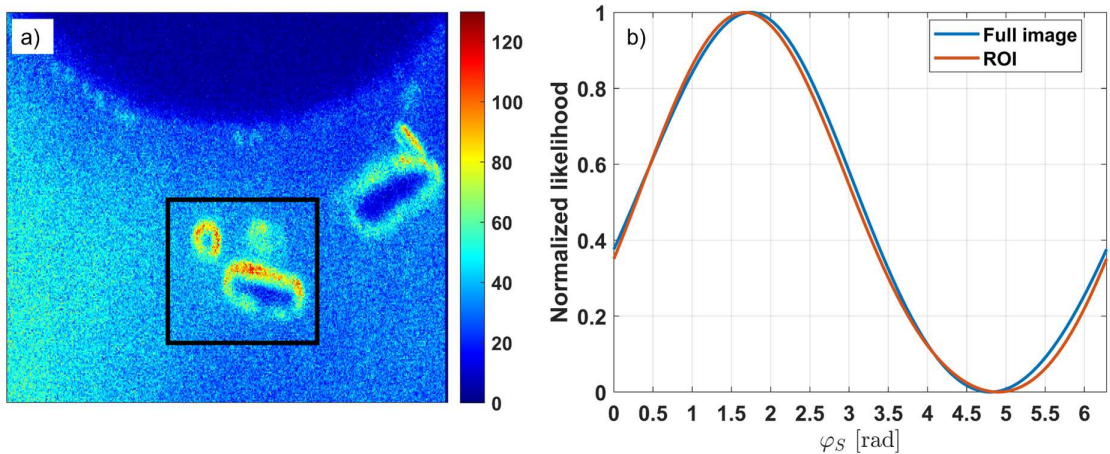

Figure 5 -Panel a) shows a sample image and the region of interest (black box) selected to perform image registration and likelihood test. In panel b) are plotted the normalized likelihood values as a function of the guessed sample phase $\varphi_s$ .

**Fundings**
E. Pa. acknowledges the funding received from the European Union's Horizon 2020 research and innovation program under the Maria Skłodowska-Curie Grant Agreement No. 860553. We acknowledge support from Laserlab Europe and from the Italian Ministry of Foreign Affairs and International Cooperation. R.H. acknowledges financial support by the German Research Foundation (DFG) through the Collaborative Research Center PolyTarget 1278, project number 316213987, subproject C04.

**Acknowledgments**
We feel that it should become a good practice to inform society about the energy cost of papers. This value is estimated to be 124 MWh. Useful discussion with F. Capotondi, A. Gianoncelli and C. Svetina are gratefully acknowledged.

**Disclosures**
The authors declare no conflicts of interest.

**Data availability**
Data underlying the results presented in this paper are not publicly available at this time but may be obtained from the authors upon reasonable request.

**Supplemental document**.
See Supplement 1 for supporting content.

**References**

1. Masters BR. Superresolution optical microscopy: the quest for enhanced resolution and contrast. Cham, Switzerland: Springer; 2020. 403 p. (Springer series in optical sciences).
2. Chen TY, Chen YT, Wang CL, Kempson IM, Lee WK, Chu YS, et al. Full-field microimaging with 8 keV X-rays achieves a spatial resolutions better than 20 nm. Opt Express. 2011 Oct 10;19(21):19919.
3. Chen YT, Lo TN, Chiu CW, Wang JY, Wang CL, Liu CJ, et al. Fabrication of high-aspect-ratio Fresnel zone plates by e-beam lithography and electroplating. J Synchrotron Radiat. 2008 Mar 1;15(2):170–5.
4. Reimer L. Transmission electron microscopy: physics of image formation and microanalysis. Berlin: Springer-Verlag; 1984.
5. Leamy HJ. Charge collection scanning electron microscopy. J Appl Phys. 1982 Jun 1;53(6):R51–80.
6. Rugar D, Hansma P. Atomic Force Microscopy. Phys Today. 1990 Oct 1;43(10):23–30.
7. Nakane T, Kotecha A, Sente A, McMullan G, Masiulis S, Brown PMGE, et al. Single-particle cryo-EM at atomic resolution. Nature. 2020 Nov 5;587(7832):152–6.
8. Kühlbrandt W. Cryo-EM enters a new era. eLife. 2014 Aug 13;3:e03678.
9. Wu M, Lander GC, Herzik MA. Sub-2 Angstrom resolution structure determination using single-particle cryo-EM at 200 keV. J Struct Biol X. 2020;4:100020.
10. Di Francia GT. Super-gain antennas and optical resolving power. Il Nuovo Cimento. 1952 Mar;9(S3):426–38.
11. Lelek M, Gyparaki MT, Beliu G, Schueder F, Griffié J, Manley S, et al. Single-molecule localization microscopy. Nat Rev Methods Primer. 2021 Jun 3;1(1):39.
12. Reinhardt SCM, Masullo LA, Baudrexel I, Steen PR, Kowalewski R, Eklund AS, et al. Ångström-resolution fluorescence microscopy. Nature. 2023 May 25;617(7962):711–6.
13. Hell SW, Wichmann J. Breaking the diffraction resolution limit by stimulated emission: stimulated-emission-depletion fluorescence microscopy. Opt Lett. 1994 Jun 1;19(11):780.
14. Gustafsson MGL, Shao L, Carlton PM, Wang CJR, Golubovskaya IN, Cande WZ, et al. Three-Dimensional Resolution Doubling in Wide-Field Fluorescence Microscopy by Structured Illumination. Biophys J. 2008 Jun;94(12):4957–70.
15. Gustafsson MGL. Surpassing the lateral resolution limit by a factor of two using structured illumination microscopy. SHORT COMMUNICATION. J Microsc. 2000

May;198(2):82–7.
16. Heintzmann R, Cremer CG. Laterally modulated excitation microscopy: improvement of resolution by using a diffraction grating. In: Bigio IJ, Schneckenburger H, Slavik J, Svanberg K, Viallet PM, editors. Stockholm, Sweden; 1999 [cited 2023 Oct 13]. p. 185–96. Available from: http://proceedings.spiedigitallibrary.org/proceeding.aspx?articleid=972650
17. Heintzmann R, Huser T. Super-Resolution Structured Illumination Microscopy. Chem Rev. 2017 Dec 13;117(23):13890–908.
18. Günther B, Hehn L, Jud C, Hipp A, Dierolf M, Pfeiffer F. Full-field structured-illumination super-resolution X-ray transmission microscopy. Nat Commun. 2019 Jun 7;10(1):2494.
19. Chergui M, Beye M, Mukamel S, Svetina C, Masciovecchio C. Progress and prospects in nonlinear extreme-ultraviolet and X-ray optics and spectroscopy. Nat Rev Phys. 2023 Sep 25;5(10):578–96.
20. Mincigrucci R, Foglia L, Naumenko D, Pedersoli E, Simoncig A, Cucini R, et al. Advances in instrumentation for FEL-based four-wave-mixing experiments. Nucl Instrum Methods Phys Res Sect Accel Spectrometers Detect Assoc Equip. 2018 Nov;907:132–48.
21. Foglia L, Mincigrucci R, Maznev AA, Baldi G, Capotondi F, Caporaletti F, et al. Extreme ultraviolet transient gratings: A tool for nanoscale photoacoustics. Photoacoustics. 2023 Feb;29:100453.
22. Nikl M, Mihokova E, Laguta V, Pejchal J, Baccaro S, Vedda A. Radiation damage processes in complex-oxide scintillators. In: Juha L, Sobierajski RH, Wabnitz H, editors. Prague, Czech Republic; 2007 [cited 2023 Oct 14]. p. 65860E. Available from: http://proceedings.spiedigitallibrary.org/proceeding.aspx?doi=10.1117/12.724737
23. Novotný P, Linhart V. Radiation damage study of thin YAG:Ce scintillator using low-energy protons. J Instrum. 2017 Jul 24;12(07):P07021–P07021.
24. Zhu R yuan. Radiation damage in scintillating crystals. Nucl Instrum Methods Phys Res Sect Accel Spectrometers Detect Assoc Equip. 1998 Aug;413(2–3):297–311.
25. Lider VV. X-ray crystal interferometers. Phys-Uspekhi. 2014 Nov 30;57(11):1099–117.
26. Nicolas C, Miron C. Lifetime broadening of core-excited and -ionized states. J Electron Spectrosc Relat Phenom. 2012 Sep;185(8–9):267–72.
27. Sanchez-Cano C, Romero-Canelón I, Yang Y, Hands-Portman IJ, Bohic S, Cloetens P, et al. Synchrotron X-Ray Fluorescence Nanoprobe Reveals Target Sites for Organo-Osmium Complex in Human Ovarian Cancer Cells. Chem – Eur J. 2017 Feb 21;23(11):2512–6.
28. Janssens K, Vittiglio G, Deraedt I, Aerts A, Vekemans B, Vincze L, et al. Use of microscopic XRF for non-destructive analysis in art and archaeometry. X-Ray Spectrom. 2000 Jan;29(1):73–91.
29. Matsuyama S, Yamada J, Kohmura Y, Yabashi M, Ishikawa T, Yamauchi K. Full-field X-ray fluorescence microscope based on total-reflection advanced Kirkpatrick–Baez mirror optics. Opt Express. 2019 Jun 24;27(13):18318.
30. Pidaparthy S, Rodrigues MTF, Zuo JM, Abraham DP. Increased Disorder at Graphite Particle Edges Revealed by Multi-length Scale Characterization of Anodes from Fast-Charged Lithium-Ion Cells. J Electrochem Soc. 2021 Oct 1;168(10):100509.
31. Diesen E, Wang HY, Schreck S, Weston M, Ogasawara H, LaRue J, et al. Ultrafast Adsorbate Excitation Probed with Subpicosecond-Resolution X-Ray Absorption Spectroscopy. Phys Rev Lett. 2021 Jun 29;127(1):016802.
32. LaRue J, Liu B, Rodrigues GLS, Liu C, Garrido Torres JA, Schreck S, et al. Symmetry-resolved CO desorption and oxidation dynamics on O/Ru(0001) probed at the C K-edge by ultrafast x-ray spectroscopy. J Chem Phys. 2022 Oct 28;157(16):164705.
33. Campbell CT, Ertl G, Kuipers H, Segner J. A molecular beam study of the catalytic oxidation of CO on a Pt(111) surface. J Chem Phys. 1980 Dec 1;73(11):5862–73.
34. Kecili R, Hussain CM. Mechanism of Adsorption on Nanomaterials. In: Nanomaterials in Chromatography [Internet]. Elsevier; 2018 [cited 2023 Aug 4]. p. 89–115. Available from: https://linkinghub.elsevier.com/retrieve/pii/B9780128127926000042

35. Chen X, Zhong S, Hou Y, Cao R, Wang W, Li D, et al. Superresolution structured illumination microscopy reconstruction algorithms: a review. Light Sci Appl. 2023 Jul 12;12(1):172.
36. Ströhl F, Kaminski CF. A joint Richardson—Lucy deconvolution algorithm for the reconstruction of multifocal structured illumination microscopy data. Methods Appl Fluoresc. 2015 Jan 16;3(1):014002.
37. Chakrova N, Rieger B, Stallinga S. Deconvolution methods for structured illumination microscopy. J Opt Soc Am A. 2016 Jul 1;33(7):B12.